\documentclass[prl,superscriptaddress,twocolumn]{revtex4}
\usepackage{graphicx} 
\def\@biblabel#1{{#1.}} 

\def\be{\begin{equation}} \def\ee{\end{equation}}

\newcommand{\ket}[1]{\mbox{$|#1\rangle$}}
\newcommand{\bra}[1]{\mbox{$\langle#1|$}}
\newcommand{\erase}[1]{}
\newcommand{\e}{\mathrm{e}} 
\renewcommand{\i}{\mathrm{i}} 

\begin{document}

\title{Experimental implementation of an adiabatic quantum optimization algorithm}
\author{Matthias Steffen} 
\email{msteffen@snowmass.stanford.edu}
\affiliation{Center for Bits and Atoms - MIT, Cambridge, Massachusetts 02139}
\affiliation{Solid State and Photonics Laboratory, Stanford University, Stanford, California
94305-4075}
\author{Wim van Dam}
\affiliation{HP Labs, Palo Alto, California 94304-1126}
\affiliation{MSRI, Berkeley, California  94720-5070}
\author{Tad Hogg}
\affiliation{HP Labs, Palo Alto, California 94304-1126}
\author{Greg Breyta}
\affiliation{IBM Almaden Research Center, San Jose, California 95120}
\author{Isaac Chuang}
\affiliation{Center for Bits and Atoms - MIT, Cambridge, Massachusetts 02139}

\begin{abstract}
  We report the realization of a nuclear magnetic resonance computer
  with three quantum bits that simulates an adiabatic quantum
  optimization algorithm.  Adiabatic quantum algorithms offer new
  insight into how quantum resources can be used to solve hard
  problems. This experiment uses a particularly well suited three
  quantum bit molecule and was made possible by introducing a
  technique that encodes general instances of the given optimization
  problem into an easily applicable Hamiltonian. Our results indicate
  an optimal run time of the adiabatic algorithm that agrees well with
  the prediction of a simple decoherence model.
\end{abstract}

\maketitle

Since the discovery of Shor's\cite{Shor94a} and Grover's
\cite{Grover97a} algorithms, the quest of finding new quantum
algorithms proved a formidable challenge. Recently however, a novel
algorithm was proposed, using adiabatic evolution
\cite{Farhi00a,Hogg00a}. Despite the uncertainty in its scaling
behavior, this algorithm remains a remarkable discovery because it
offers new insights into the potential usefulness of quantum
resources for computational tasks.

Experimental realizations of quantum algorithms in the past
demonstrated Grover's search algorithm \cite{Jones98a,Vandersypen00a},
the Deutsch-Jozsa algorithm \cite{Chuang98b,Marx00a,Jones98b},
order-finding \cite{Vandersypen00b}, and Shor's algorithm
\cite{Vandersypen01a}. Recently, Hogg's algorithm was implemented
using only one computational step ~\cite{Peng02a}, however a
demonstration of an adiabatic quantum algorithm thus far has remained
beyond reach.

Here, we provide the first experimental implementation of an adiabatic
quantum optimization algorithm using three qubits and nuclear magnetic
resonance (NMR) techniques \cite{Gershenfeld97a,Cory97a}. NMR
techniques are especially attractive because several tens of qubits
may be accessible, which is precisely the range that could be crucial
in determining the scaling behavior of adiabatic quantum algorithms
\cite{Farhi01a}. Compared to earlier implementations of search
problems \cite{Jones98a,Vandersypen00a}, this experiment is a full
implementation of a true optimization problem, which does not require
a black box function or ancilla bits.

This experiment was made possible by overcoming two experimental
challenges. First, an adiabatic evolution requires a smoothly varying
Hamiltonian over time, but the terms of the available Hamiltonian in
our system cannot be smoothly varied and may even have fixed values.
We developed a method to approximately smoothly vary a Hamiltonian
despite the given restrictions by extending NMR average Hamiltonian
techniques \cite{Rhim70a}. Second, general instances of
the optimization algorithm may require the application of Hamiltonians
that are not easily accessible. We developed methods to implement
general instances of a well known classical NP-complete optimization
problem given a fixed natural system Hamiltonian.

We provide a concrete procedure detailing these methods. We then apply
the results to our optimization problem which is known as Maximum Cut
or {\sc maxcut} \cite{Garey76a}.  Our experimental results indicate
there exists an optimal total running time which can be predicted
using a decoherence model that is based on independent stochastic
relaxation of the spins.

The evolution of a quantum state during an adiabatic quantum algorithm
is determined by a slowly varying, time-dependent Hamiltonian. Suppose
we are given some time-dependent Hamiltonian $H(t)$ where $0 \leq t
\leq T$, and at $t=0$ we start in the ground state of $H(0)$. By
varying $H(t)$ slowly, the quantum system remains in the ground state
of $H(t)$ for all $0 \leq t \leq T$ provided the lowest two energy
eigenvalues of $H(t)$ are never degenerate \cite{Messiah76a}. Now
suppose we can encode an optimization problem into $H(T)$. Then the
state of the quantum system at time $t=T$ represents the solution to
the optimization problem \cite{Farhi00a}. The total run-time $T$ of
the adiabatic algorithm scales as $g^{-2}_{min}$ where $g_{min}$ is
the minimum separation between the lowest two energy eigenvalues of
$H(t)$ \cite{Farhi00a,VanDam02a}. It is the scaling behavior of
$g_{min}$ that will ultimately determine the success of adiabatic
quantum algorithms. Classical simulations of this scaling behavior are
hard due to the exponentially growing size of Hilbert space. In
contrast, sufficiently large quantum computers could simulate this
behavior efficiently.

Smoothly varying some time-dependent Hamiltonian appears
straightforward but it contrasts with the traditional picture of
discrete unitary operations including fault tolerant quantum circuit
constructions \cite{Aharonov97a}. Fortunately, we can approximate a
smoothly varying Hamiltonian using methods of quantum simulations
\cite{Trotter58a} and recast adiabatic evolution in terms of unitary
operations.

Discretizing a continuous Hamiltonian is a straightforward process
and changes the run time $T$ of the adiabatic algorithm only
polynomially \cite{VanDam02a}. For simplicity, let the discrete time
Hamiltonian $H[m]$ be a linear interpolation from some beginning
Hamiltonian $H[0]=H_b$ to some final problem Hamiltonian $H[M]=H_p$
such that $H[m] = (m/M)H_p + (1-m/M)H_b$.  The unitary evolution of the
discrete algorithm can be written as: 
\be
U = \prod_m U_m = \prod_m \e^{-\i((1-m/M)H_b + (m/M)H_p)\Delta t}
\label{discreteideal}
\ee 
where $\Delta t = T/(M+1)$, and $M+1$ is the total number of
discretization steps. The adiabatic limit is achieved when both $T,M
\rightarrow \infty$ and $\Delta t \rightarrow 0$.

Full control over the strength of $H_b$ and $H_p$ is needed to
implement Eq.(\ref{discreteideal}). However, this may not necessarily
be a realistic experimental assumption. We will next show how the
discrete time adiabatic algorithm can still be implemented when $H_b$
and $H_p$ cannot both be applied simultaneously {\it and} when they
are both fixed in strength.

When both $H_b$ and $H_p$ are fixed, we can approximate $U_m$ to
second order by using the Trotter formula $\mathrm{exp}((A+B)\Delta t)
= \mathrm{exp}(A\Delta t/2)\mathrm{exp}(B\Delta t)\mathrm{exp}(A\Delta
t/2) + {\cal O}(\Delta t^2)$ \cite{Trotter58a}. Higher order
approximations can be constructed if more accuracy is required.

Now suppose $H_b$ and $H_p$ are both constant. Since any unitary
matrix is generated by an action $-\i H \Delta t$, we can increase the
effect of a constant Hamiltonian $H$ by lengthening the time $\Delta
t$. Thus, we can implicitly increase the strength of $H_b$ and $H_p$
even when they are constant by simply increasing the time during which
they are applied.

\begin{figure}[htbp]
\begin{center}
\fbox{\includegraphics*[trim=200 170 170 80,angle=90,width=3.25in]{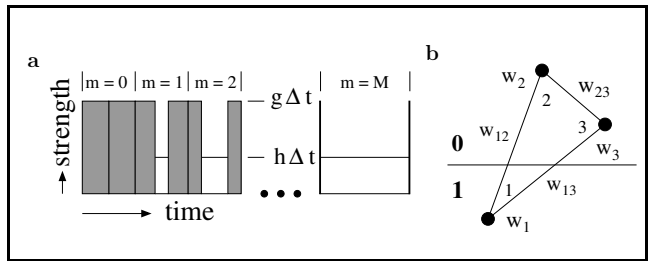}} 
\end{center}
\caption{(a) Illustration of Eq.(\ref{discreteexp}). 
The shaded and clear boxes denote the strength and duration of the 
Hamiltonians $H_b$ and $H_p$ respectively. (b) Illustration of a 
graph consisting of 3 nodes and 3 edges. The edges carry weights 
$w_{12}$, $w_{13}$, and $w_{23}$.  When $\mathrm{min}(w_{ij}) = w_{23}$ 
as indicated by the length of the edges, the {\sc maxcut} corresponds 
to the drawn cut. The solution is therefore $s=100$ and also $s=011$ 
due to symmetry. This symmetry can be broken by assigning the weights 
$w_1$, $w_2$, and $w_3$ to the nodes.}
\label{fig:maxcut}
\end{figure}

This technique also allows cases when the accessible Hamiltonians are
not of the required strength, for example when we are given
$H_b'=gH_b$ and $H_p'=hH_p$ but still wish to implement $H_b$ and
$H_p$. Using all of the described techniques, we can now write $U_m$
as:
\be
U_m \approx \e^{-\i H_b'[(1-m/M)\Delta t/2g]}\circ \e^{-\i H_p'[(m/M)\Delta t/h]}
\label{discreteexp}
\ee
where $A \circ B = ABA$. Each discretization step is of length
$(1-m/M)\Delta t/g + (m/M)\Delta t/h$, which is not constant 
when $g \neq h$. As an illustration consider Fig.~\ref{fig:maxcut}a. 

In this experiment we choose $\Delta t = T/(M+1)$ to be constant as we
vary the number of discretization steps $M+1$. This way, the total run
time $T$ increases with $M+1$, allowing us to test the behavior of the
algorithm when approaching one of the conditions for the adiabatic
limit.  Even when the discrete approximation is not close to the
adiabatic limit, the implemented algorithm can often find solutions
using relatively few steps but lacks the guaranteed performance of the
adiabatic theorem \cite{Hogg00b}.

Adiabatic evolution has been proposed to solve general optimization
problems, including NP-complete ones.  In this general setting, the
algorithm can depend on the existence of a black box function or the
usage of large amounts of workspace.  Our goal here is to optimize a
hard natural problem in a way that avoids these difficulties. We will
first describe which problem we chose and later on explain why it does
not require ancilla qubits.

We found the {\sc maxcut} problem to be a well-suited problem to
demonstrate an adiabatic quantum algorithm because it allows a variety
of interesting test cases. It also has applications in the study of
spin glasses \cite{Barahona85a} and VLSI design \cite{Sarrafzadeh98a},
among others.  The decision variant of the {\sc maxcut} problem is
part of the core NP-complete problems \cite{Garey76a}, and even the
approximation within a factor of 1.0624 of the perfect solution is
NP-complete \cite{Ausiello99a}.

The {\sc maxcut} problem can be understood as follows. A {\em cut} is
defined as the partitioning of an undirected $n$-node graph with edge
weights into two sets.  We define the payoff as the sum of weights of
edges crossing the cut.  The maximum cut is a cut that maximizes this
payoff. By assigning either $s_i = 0$ or $s_i = 1$ to each node $i$,
depending on its location with respect to the cut, the {\sc maxcut}
problem can be restated as finding the $n$ bit number $s$ that
maximizes the payoff.  An extension of the {\sc maxcut} problem is to
let the nodes themselves carry weights, which can be regarded as the
nodes having a preference on their location. As an illustration
consider a graph with three nodes as drawn in Fig.~\ref{fig:maxcut}b.

The payoff as a function of the cut defined by $s$ is given by
\be 
P(s) = \sum_i w_i s_i + \sum_{i,j} s_i(1-s_j)w_{ij}
\label{maxcutcost1}
\ee 
where $w_{ij}$ are the edge weights, $w_i$ denotes the preference
of the nodes to be on the $1$ side of the cut, and $s_i$ is the value
of the $i$-th bit of $s$, for $0 \leq s \leq 2^n-1$.

The smallest meaningful test case of the {\sc maxcut} problem requires
$3$ nodes and admits a variety of interesting cases by varying $w_i$
and $w_{ij}$. We aimed at two goals when choosing a representative set
of weights. First, we wanted the minimum energy gap $g_{min}$ to be
smaller than the one for a $3$-qubit adiabatic Grover search. Second,
we wanted a resulting energy landscape with both a global and local
maximum such that a greedy classical search would incorrectly find the
local maximum half the time~\cite{greedydef}. These goals are met by
the choice $w_1 = w_2 = w_3 = 2$, $w_{12}=2$, $w_{13}=1$, $w_{23}=3$.
The payoff function for this set of weights is $P(s) = [0\ 6\ 7\ 7\ 5\ 
{\bf 9} \ 8\ 6]$ where $s$ = [000 001 010 011 100 101 110 111]. The
global maximum lies at $s=101$ so the answer on the quantum computer
following measurement should be $\ket{101}$, and not at the local
maximum $s=110$.

In the quantum setting, this payoff function $P(s)$ can be encoded into
the Hamiltonian $H_p$ by rewriting Eq.(\ref{maxcutcost1}) 
using Pauli matrices: 
\be 
H_p = \sum_i w_i(I-\sigma_{zi})/2 + \sum_{i<j} w_{ij}(I-\sigma_{zi}\sigma_{zj})/2
\label{maxcutcost2}
\ee 
where $I$ is the $2^n$x$2^n$ identity matrix and $\sigma_{zi}$ is the
Pauli $Z$ matrix on spin $i$. The identity matrices in the equation
above only lead to an overall phase which cannot be observed, and
hence they can be ignored. The diagonal values of
Eq.(\ref{maxcutcost2}) are equal to $P(s)$. Because of the direct
encoding of $P(s)$ into $H_p$ no black box function or ancilla qubits
are required, which makes this a full implementation of an
optimization problem.

Similar to Eq.(\ref{maxcutcost2}), the natural Hamiltonian of $n$
weakly coupled spin-1/2 nuclei subject to a static magnetic field
$B_0$ is well approximated by \cite{Ernst94a} 
\be 
{\cal H} = -\sum_i \omega_i\sigma_{zi}/2 + \sum_{i<j} \pi J_{ij}\sigma_{zi}\sigma_{zj}/2 + {\cal H}_{env}
\label{hamilt}
\ee 
where the first term represents the Larmor precession of each spin
$i$ about $-B_0$, and $\omega_i$ is its Larmor frequency.  The second
term describes the scalar spin-spin coupling of strength $J_{ij}$
between spins $i$ and $j$. The last term represents coupling to the
environment, causing decoherence. Note the resemblances between ${\cal
  H}$ and $H_p$.
 
Despite the similarities, the spin-spin couplings of Eq.(\ref{hamilt})
are generally different from a randomly chosen set of weights.
Therefore, we require a procedure to turn the fixed $J_{ij}$ into any
specified weights $w_{ij}$. This is achieved using refocusing schemes
that are typically used to turn on only one of the couplings while
turning all others off \cite{Ernst94a}. 

\begin{figure}[htbp]
\begin{center}
\fbox{\includegraphics*[angle=90,width=3.2in,trim=250 240 220 50]{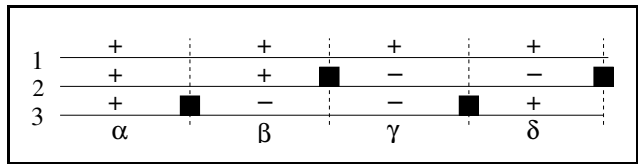}}
\end{center}
\caption{Refocusing scheme to effectively change $J_{ij}$ into $w_{ij}$. 
The horizontal lines denote qubits $1$, $2$, and $3$ and time goes from 
left to right. The black rectangles represent $180^\circ$ rotations. 
The delay segments are of length $\alpha$, $\beta$, $\gamma$, and $\delta$. 
When all segments are of equal length, all couplings are effectively turned 
off \cite{Leung00a} because $\sigma_{xi} \e^{-\i \sigma_{zi}\sigma_{zj}t} 
\sigma_{xi} = \e^{\i \sigma_{zi}\sigma_{zj}t}$. In our experiment, 
$\alpha=0.42$ ms, $\beta=0$ ms, $\gamma=4$ ms, and $\delta=2.9$ ms in 
the last slice $M+1$. The RF pulses that implement $Hb'$ perform 
$33.75^\circ$ rotations on the qubits in the first slice.}
\label{fig:refocus}
\end{figure}

We have modified a refocusing scheme to effectively change the
couplings to any arbitrary value. Consider the pulse sequence drawn in
Fig.~\ref{fig:refocus}. Based on this scheme, we can derive the
under-constrained system $(\alpha+\beta-\gamma-\delta)J_{12}=w_{12}$,
$(\alpha-\beta-\gamma+\delta)J_{13}=w_{13}$, and
$(\alpha-\beta+\gamma-\delta)J_{23}=w_{23}$, which can be solved for
positive $\alpha$, $\beta$, $\gamma$, and $\delta$ such that $J_{ij} \rightarrow w_{ij}$.

The single weights $w_i$ are implemented by introducing a reference
frame for each spin $i$ which rotates about $-B_0$ at frequency
$(\omega_i - w_i)/2$. In order to apply the single qubit rotations of
our refocusing scheme on resonance, we apply the reference frequency
shift only during the delay segment $\alpha$, which we can always
choose to be a positive value. Thus, $H_p$ is implemented by applying
the refocusing scheme from Fig.~\ref{fig:refocus} while going
off-resonance during the delay segment $\alpha$.
 
A full implementation of an adiabatic algorithm also requires a proper
choice of $H_b$. We choose $H_b = \sum_i \sigma_{xi}$ for several
reasons. First, its highest two excited states are non-degenerate.
Second, it can be easily generated using single qubit rotations, and
third, its highest excited state is created from a pure state with all
qubits in the $\ket{0}$ state by applying a Hadamard gate on all
qubits (we require the initial state to be the {\em highest excited}
state of $H_b$ because we are optimizing for the {\em maximum} value
of $H_p$).

The full adiabatic quantum algorithm is now implemented by first
creating the highest excited state of $H_b$. We then apply $M+1$
unitary matrices as given by Eq.(\ref{discreteexp}) and illustrated
by Fig.\ref{fig:maxcut}a. Accordingly, from slice to slice, we
decrease the time during which $H_b$ is active while increasing the
time during which $H_p$ is active. Finally, we measure the quantum
system and read-out the answer.

We selected $^{13}$C-labeled CHFBr$_2$ for our experiments. The
Hamiltonian of the $^1$H-$^{19}$F-$^{13}$C system is of the form of
Eq.(\ref{hamilt}) with measured couplings $J_{HC}=224$ Hz, $J_{HF}=50$
Hz, and $J_{FC}=-311$ Hz. The interaction with the Br nuclei is
averaged out, contributing only to ${\cal H}_{env}$. Experiments were
carried out at MIT using an 11.7 Tesla Oxford Instruments magnet and a
Varian Unity Inova spectrometer with a triple resonance (H-F-X) probe
from Nalorac.

The experiments were performed at room temperature at which the
thermal equilibrium state is highly mixed and cannot be turned into
the required initial state by just unitary transforms. We thus first
created an approximate effective pure state as in
ref.~\onlinecite{Vandersypen00a} by summing over three temporal
labeling experiments.

In our experiments, we actually implemented $0.5H_p$ and $0.5887H_b$
instead of $H_p$ and $H_b$.  This ensures that the error due to the
2nd order Trotter approximation is bounded by $\sqrt{\sum_i
  |\epsilon_i|^2} < 0.0356$ where $\epsilon_i$ is the contribution of
the $i$-th undesired Pauli matrix. We also choose $g$ so the applied
RF field does not heat the sample, and $g \gg h$ so $J_{ij}$ can be
ignored when applying $H_b$. All of these choices result in a total
experimental time that is within the shortest $T_2$ decoherence time
\cite{Vandersypen00a}. We reconstructed the traceless deviation
density matrices upon completion of the experiments using quantum
state tomography \cite{Vandersypen00a}.

\begin{figure}[htbp]
\begin{center}
\fbox{\includegraphics*[angle=90,width=3.2in,trim = 180 185 175 80]{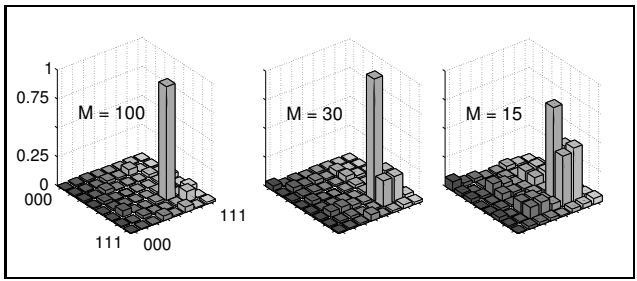}}
\end{center}
\caption{Plot of the absolute value of the deviation density matrix for 
$M=100$ ($T=374$ ms), $M=30$ ($T=115$ ms), and $M=15$ ($T=59.2$ ms), 
adjusted by an identity portion such that the minimum diagonal value 
equals zero. The scale is arbitrary but the same for each plot.}
\label{fig:devdenadia}
\end{figure}

We executed this algorithm for several $M$ (with $w_i$ and $w_{ij}$ as
listed above Eq.(\ref{maxcutcost2})). Since we chose $\Delta t$ to be
constant, this meant increasing the run-time $T$ of the algorithm. The
reconstructed deviation density matrices are shown in
Fig.~\ref{fig:devdenadia}. The plots clearly display the expected pure
state $\ket{101}$.  The local maximum at $s = 110$ has a decreasingly
small probability of being measured for increasing $M$. Simulations
using Eq.(\ref{discreteexp}) show that this optimization algorithm
performs better for increasing $M$. We wanted to verify whether this
is indeed true experimentally.

\begin{figure}[htbp]
\begin{center}
\fbox{\includegraphics*[angle=90,width=3.2in,trim = 200 195 200 100]{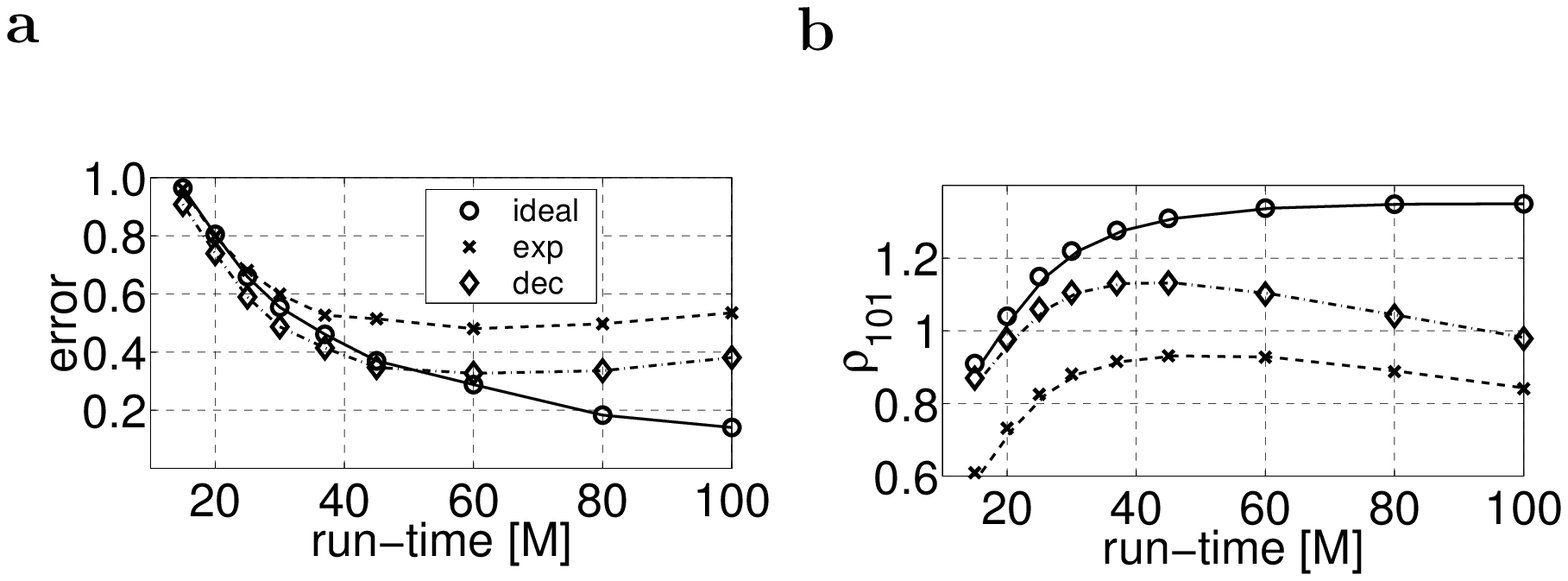}}
\end{center}
\caption{Experimental performance of the adiabatic algorithm. 
(a) Plot of the error as a function of $M$. The error measure is the 
trace distance $D(\rho,\sigma) = |\rho-\sigma|/2$ where $\sigma$ is the 
traceless deviation density matrix for $M=400$, approximating 
$M \rightarrow \infty$, and $\rho$ equals the ideal expected ($\circ$), 
the experimentally obtained ($\times$), or the ideal expected traceless 
deviation density matrix with decoherence effects ($\diamond$) 
\cite{Vandersypen01a}. The minimum error occurs at about $M=60$ indicating 
an optimal run-time of the algorithm. 
(b) A similar observation can be made when plotting $\ket{101}\bra{101}$
as a function of $M$.}
\label{fig:errorplot}
\end{figure}

For this purpose, we estimate the error of our obtained deviation
density matrices compared with the ideal case of $M=\infty$.
Fig.~\ref{fig:errorplot}a plots the trace distance as a function of
$M$, using the same arbitrary scale as in Fig.~\ref{fig:devdenadia}. From
the plot, we observe there exists an optimal run-time of the
algorithm, corresponding to $0.226$ seconds in our experiment. This
optimal run time is in good agreement with the prediction of a previously
developed simple decoherence model \cite{Vandersypen01a}.  Predicting
the impact of decoherence has already provided invaluable insight into
estimating errors in previous experiments \cite{Vandersypen01a}, and
we believe continued effort towards understanding decoherence will
greatly benefit experimental investigations of quantum systems.

In conclusion, we have provided the first experimental demonstration
of an adiabatic quantum optimization algorithm. We show a concrete
procedure turning a continuous time adiabatic quantum algorithm into a
discrete time version, even when certain restrictions apply to the
accessible Hamiltonians. Our results indicate that there exists an
optimal run-time of the algorithm which can be roughly predicted using
a simple decoherence model. We believe this implementation opens the
door to a variety of interesting experimental demonstrations and
investigations of adiabatic quantum algorithms.

\begin{acknowledgments}
  We wish to thank A. Childs, A. Landahl, and E. Farhi for useful
  discussions. This work was supported by the NSF grant CCR-0122419,
  the DARPA QuIST program, the HP/MSRI postdoc fellowship.
\end{acknowledgments}

\bibliography{mybib}

\begin{thebibliography}{28}
\bibitem[{gre()}]{greedydef}
\bibinfo{note}{A greedy search is done by first choosing a random node
  configuration $s$, and then repeatedly moving to a new configuration $s'$
  which differs from the previous configuration by only one node value $s'_i$
  and which also has the highest payoff, until the payoff is maximized.}
\end{thebibliography}
\end{document}